\documentstyle[12pt,aasms4]{article}

\slugcomment{submitted to}

\lefthead{Xiangdong Shi}
\righthead{Peculiar Hubble Flows in Our Local Universe}

\begin{document}

\title{Peculiar Hubble Flows in Our Local Universe}

\author{Xiangdong Shi}
\affil{shi@doheny.ucsd.edu\\
Department of Physics, University of California, La Jolla, 
CA 92093-0354}

\begin{abstract}
A formalism that simultaneously searches for the monopolar
and dipolar peculiar velocities is presented.  The formalism is
applied to (1) the Mark III catalogue, (2) Lauer and Postman's Abell cluster
catalogue, and (3) Riess et al.'s Type Ia supernova sample. 
The emphasis is drawn to the monopolar peculiar velocities, i.e.,
peculiar Hubble flows, within these samples.  
The samples show inconsistent peculiar Hubble flows within a depth of
$\sim 60h^{-1}$ Mpc. Beyond a depth of $\sim 80h^{-1}$ Mpc, the
Hubble flows of all samples converge to the global Hubble flow
to better than 10$\%$ at the $2\sigma$ level.
The results are compared with theoretical predictions.  They at
face value disfavor models predicting smaller peculiar velocities such as
the tilted Cold Dark Matter model.  Limitations of the catalogues
are discussed, so are ways to improve the catalogues so that
an accurate map of Hubble flows in our local universe can be drawn
and be compared with theoretical predictions.
\end{abstract}


\section{Introduction}

Peculiar Hubble flows (monopolar deviations from a global Hubble flow)
are as important as bulk motions (dipolar deviations from a global
Hubble flow) in reflecting the underlying density fluctuation of the universe. 
But they have not received the same attention as bulk motions have, because
the uncertainty of our knowledge about the true Hubble constant $H_0$
is often larger than the expected peculiar Hubble flows at large scales.
One can, however, investigate the variation of Hubble flows within
a sample without knowing the value of $H_0$.  Moreover,
the inferred Hubble flows from samples that extend
significantly beyond 100$h^{-1}$ Mpc should be so close
to the global Hubble flow that it is meaningful to investigate the
peculiar Hubble flows within the samples.
In doing so, the ultimate goal is to map out the variation of the
Hubble expansion as a function of depth, which is directly related
to the underlying density fluctuation (Shi, Widrow and Dursi 1996;
Turner, Cen and Ostriker 1992), and then to use the peculiar Hubble flows
detected to test theoretical models.

Here I first present a formalism that simultaneously calculates the
peculiar Hubble flows and the bulk motions, along with their errors, within a
sample.  The formalism is applied to (1) the Mark III catalogue based
on the Tully-Fisher relation as a distance indicator (Willick et al. 1996),
(2) The Abell cluster catalogue of Lauer and Postman (1994, LP thereafter)
with the brightest cluster galaxies (BCGs)
as distance indicators, and (3) the sample of 20 Type Ia
supernovae (SNe) of Riess et al. (Riess, Press, Kirshner 1996).
All three samples are deep or fairly deep: the Mark III catalogue
extends to 11,000 km/sec; the LP sample is complete up to 15,000 km/sec;
the SN sample probes the deepest, 28,000 km/sec.  
In other words, Hubble flows defined by these samples should approximate the
global Hubble flow to a fairly high degree (see table 1 and text).

The results from the three samples are presented and analyzed in section 3--5.
They are compared with theoretical predictions from four representative
models (table 1): the standard Cold Dark Matter (sCDM)
model, a tilted Cold Dark Matter (tCDM) model, a vacuum energy $\Lambda$
dominated Cold Dark Matter ($\Lambda$CDM) model, and a Cold plus Hot Dark
Matter (CHDM) model.  Finally I conclude in section 6 and comment briefly on
improving the measurements of peculiar Hubble flows and their potential
to test theoretical models.

\section{Formalism}

The peculiar Hubble flow and the bulk motion of a sample can be obtained by
maximizing the likelihood (Kaiser 1988, 1991)
\begin{equation}
L(U_i,\delta H)=\prod_q{1\over\sigma_q^2}\exp\Bigl[-{(S_q-\delta Hr_q-
                            U_i\hat r_q^i)^2\over 2\sigma_q^2}\Bigr],
\label{likelihood}
\end{equation}
where ${\bf r}_q$ ($=r_q^i{\bf\hat r}_i, i=x, y, z$)
is the position of an object in the sample,
$S_q$ its estimated line-of-sight peculiar velocity with
an uncertainty $\sigma_q$. ${\bf U}$ ($=U_i{\bf\hat r}_i, i=x, y, z$) is the
bulk motion of the sample, and $\delta H$ is its peculiar Hubble expansion
rate. Maximizing $L(U_i,\delta H)$ with respect to $U_i$ and $\delta H$ gives
\begin{equation}
U_i=(A-RB^{-1})^{-1}_{ij}\Bigl(\sum_q{S_q{\hat r}_q^j\over\sigma_q^2}-
    B^{-1}\sum_q\sum_{q^\prime}{S_qr_qr_{q^\prime}^j\over
    \sigma_q^2\sigma_{q^\prime}^2}\Bigr),
\label{bulkmotion}
\end{equation}
and
\begin{equation}
\delta H=B^{-1}\sum_q{S_qr_q-U_ir_q^i\over\sigma_q^2},
\label{pechub}
\end{equation}
where
\begin{equation}
A_{ij}=\sum_q{\hat r_q^i\hat r_q^j\over\sigma_q^2},\quad
R_{ij}=\sum_q\sum_{q^\prime}{r_q^ir_{q^\prime}^j
\over\sigma_q^2\sigma_{q^\prime}^2},\quad
B=\sum_q{r_q^2\over\sigma_q^2}.
\label{Aij}
\end{equation}
$S_q$ is related to the real peculiar velocity field ${\bf v}({\bf r})$ by
\begin{equation}
S_q=v_i({\bf r}_q)\hat r_q^i+\epsilon_q,
\label{Sq}
\end{equation}
where $\epsilon_q$ is a gaussian random variable with an uncertainty 
$\sigma_q$. Therefore,
\begin{equation}
\delta H=\delta H^{(v)}+\delta H^{(\epsilon)}.
\label{v+eps}
\end{equation}
The first term is the noise-free contribution from density fluctuations
\begin{equation}
\delta H^{(v)}=\int d^3r W^i({\bf r})v_i({\bf r}),
\label{dHv}
\end{equation}
while the second term represents the contribution from noises in data.

The window function in eq. (\ref{dHv}) is

$W^i({\bf r})=$
\begin{equation}
\hat r^iB^{-1}\Bigl\{\sum_q {r_q\over\sigma_q^2}\delta ({\bf r}-{\bf r}_q)-
    (A-RB^{-1})^{-1}_{jl}\Bigl[\sum_q{\hat r_q^j\over\sigma_q^2}
    \delta ({\bf r}-{\bf r}_q)-
    B^{-1}\sum_q\sum_{q^\prime}{r_qr_{q^\prime}^j\over
    \sigma_q^2\sigma_{q^\prime}^2}\delta ({\bf r}-{\bf r}_q)\Bigr]
    \sum_{q^{\prime\prime}}{r_{q^{\prime\prime}}^l
    \over\sigma_{q^{\prime\prime}}^2}\Bigr\}.
\end{equation}
Its Fourier transformation is
\begin{equation}
W^i({\bf k})={B^{-1}\over (2\pi)^{3/2}}
   \Bigl[\sum_q {r_q^i\over\sigma_q^2}e^{i{\bf k}\cdot{\bf r}_q}-
   (A-RB^{-1})^{-1}_{jl}\Bigl(\sum_q{\hat r_q^i\hat r_q^j\over\sigma_q^2}
   e^{i{\bf k}\cdot{\bf r}_q}-
   B^{-1}\sum_q\sum_{q^\prime}{r_q^ir_{q^\prime}^j\over
   \sigma_q^2\sigma_{q^\prime}^2}e^{i{\bf k}\cdot{\bf r}_q}\Bigr)
   \sum_{q^{\prime\prime}}{r_{q^{\prime\prime}}^l
   \over\sigma_{q^{\prime\prime}}^2}\Bigr].
\end{equation}

The expected r.m.s. peculiar Hubble flow for the sample is
\begin{equation}
\Bigl\langle\Bigl({\delta H\over H_0}\Bigr)^2\Bigr\rangle=
\Bigl\langle\Bigl({\delta H^{(v)}\over H_0}\Bigr)^2\Bigr\rangle +
\Bigl\langle\Bigl({\delta H^{(\epsilon)}\over H_0}\Bigr)^2\Bigr\rangle.
\label{sigself}
\end{equation}
In linear theory,
\begin{equation}
\Bigl\langle\Bigl({\delta H^{(v)}\over H_0}\Bigr)^2\Bigr\rangle
=\Omega_0^{1.2}\int d^3k
\bigl\vert W^i({\bf k})\hat k_i\bigr\vert ^2{P(k)\over k^2}.
\end{equation}
From eq.~(\ref{Sq}) and (\ref{v+eps})
\begin{eqnarray}
\Bigl\langle\Bigl({\delta H^{(\epsilon)}\over H_0}\Bigr)^2\Bigr\rangle
=&B^{-2}\Bigl\langle\Bigl[\sum_q{\epsilon_qr_q\over\sigma_q^2}-
    (A-RB^{-1})^{-1}_{ij}\Bigl(\sum_q{\epsilon_q{\hat r}_q^j\over\sigma_q^2}-
    B^{-1}\sum_q\sum_{q^\prime}{\epsilon_qr_qr_{q^\prime}^j
    \over\sigma_q^2\sigma_{q^\prime}^2}\Bigr)\sum_{q^{\prime\prime}}
    {r_{q^{\prime\prime}}^i\over\sigma_{q^{\prime\prime}}^2}\Bigr]\cr
 &  \Bigl[\sum_q{\epsilon_qr_q\over\sigma_q^2}-
    (A-RB^{-1})^{-1}_{lm}\Bigl(\sum_q{\epsilon_q{\hat r}_q^m\over\sigma_q^2}-
    B^{-1}\sum_q\sum_{q^\prime}{\epsilon_qr_qr_{q^\prime}^m
    \over\sigma_q^2\sigma_{q^\prime}^2}\Bigr)\sum_{q^{\prime\prime}}
    {r_{q^{\prime\prime}}^l\over\sigma_q^2}\Bigr]\Bigr\rangle\cr
=&B^{-1}+B^{-2}(A-RB^{-1})^{-1}_{il}R_{il}.\quad\quad\quad\quad\quad\quad
  \quad\quad\quad\quad\quad\quad\quad\quad\quad\quad\quad\quad
\label{noiseself}
\end{eqnarray}
Similarly,
\begin{equation}
\Bigl\langle U_i^{(\epsilon)} U_j^{(\epsilon)}\Bigr\rangle
=(A-RB^{-1})^{-1}_{ij}.
\label{Unoise}
\end{equation}

If the true Hubble constant is uncertain and one wishes to
investigate the variation of the Hubble flow within the sample,
it is more appropriate to calculate the r.m.s. deviation
of the Hubble flow of a subsample (with a rate $H_2$) from the Hubble flow
of the whole sample (with a rate $H_1$). Thus
$$\Bigl\langle\Bigl({H_1-H_2\over H_1}\Bigr)^2\Bigr\rangle
=\Bigl\langle\Bigl({\delta H_1+H_0-\delta H_2-H_0\over \delta H_1+H_0}\Bigr)^2
 \Bigr\rangle
\approx\Bigl\langle\Bigl({\delta H_1\over H_0}\Bigr)^2
+\Bigl({\delta H_2\over H_0}\Bigr)^2-2\Bigl({\delta H_1\over H_0}\Bigr)
 \Bigl({\delta H_2\over H_0}\Bigr)\Bigr\rangle$$
\begin{equation}
=\Bigl\langle\Bigl({\delta H_1^{(v)}\over H_0}\Bigr)^2
+\Bigl({\delta H_2^{(v)}\over H_0}\Bigr)^2-2\Bigl({\delta H_1^{(v)}\over H_0}
 \Bigr)\Bigl({\delta H_2^{(v)}\over H_0}\Bigr)\Bigr\rangle
+\Bigl\langle\Bigl({\delta H_1^{(\epsilon)}\over H_0}\Bigr)^2
+\Bigl({\delta H_2^{(\epsilon)}\over H_0}\Bigr)^2
-2\Bigl({\delta H_1^{(\epsilon)}\over H_0}\Bigr)
\Bigl({\delta H_2^{(\epsilon)}\over H_0}\Bigr)\Bigr\rangle,
\label{variation}
\end{equation}
in which
\begin{equation}
\Bigl\langle\Bigl({\delta H_1^{(v)}\over H_0}
\Bigr)\Bigl({\delta H_2^{(v)}\over H_0}\Bigr)\Bigr\rangle
=\Omega_0^{1.2}\int d^3k
\bigl\vert W^i_1({\bf k})\hat k_i{W^j_2}^*({\bf k})\hat k_j
\bigr\vert {P(k)\over k^2}
\label{sigcross}
\end{equation}
and
\begin{equation}
\Bigl\langle\Bigl({\delta H_1^{(\epsilon)}\over H_0}\Bigr)
\Bigl({\delta H_2^{(\epsilon)}\over H_0}\Bigr)\Bigr\rangle=
\Bigl\langle\Bigl({\delta H_1^{(\epsilon)}\over H_0}\Bigr)^2\Bigr\rangle
.
\label{noisecross}
\end{equation}
Subscript ``1'' refers to quantities of the entire sample, and subscript ``2''
refers to those of the subsample.

\section{Peculiar Hubble Flows in the Mark III catalogue}

The Mark III catalogue (Willick et al. 1996) compiled the distances and
redshifts to 
about 3000 galaxies, based on a template Tully-Fisher relation from
the 36 clusters of Han \& Mould (HM sample thereafter; Mould et al. 1991;
Han and Mould 1992; Mould et al. 1993).  Therefore
the underlying Hubble expansion of the Mark III catalogue is defined by
the HM sample.  Table 1 lists the expected 1$\sigma$ deviation of this
underlying Hubble expansion rate $H_1$ from a global Hubble constant $H_0$ in
various models, which is typically 2$\%$ to $3\%$.  (In this and
all the following calculations, clusters are assumed to have
a gaussian random motion with a one-dimension dispersion of 300 km/sec, and
all bulk motions are in reference to the Cosmic Microwave Background.)
Figure 1(a) shows the variation of Hubble flows in the HM sample with 2$\sigma$
errorbars calculated from eqs.~(\ref{bulkmotion}) to (\ref{noisecross}) vs.
the depth $R$ of subsamples.  Each subsample is
defined to be the $N$ clusters closest to us, and $R$ is the distance to
the furthest cluster in the subsample.
The errors of the variation are estimated from
the second term of eq.~(\ref{variation}).  In particular, they include the
contribution from uncertainties in determining bulk motions.

The figure shows significant positive variation within 40 to
$60h^{-1}$ Mpc, translating into significant peculiar Hubble flows
$H_2-H_0$ given the smallness of $H_1-H_0$.
In figure 1(b) the variation is compared with 2$\sigma$
expectations (with noise included) from theoretical models.
Clearly the positive $(H_2-H_1)/H_1$ found exceed the
2$\sigma$ expectation from sCDM within $\sim 50h^{-1}$ Mpc, and exceed
the 2$\sigma$ expectations from CHDM, tCDM and $\Lambda$CDM within 
a wider range of subsamples.  The variation at scales beyond
$\sim 60h^{-1}$ Mpc, however, is perfectly consistent with all models at
2$\sigma$.  Not only the variation of Hubble
flows in the HM sample is extremely large, its bulk flow is also extremely
large, at ($740\pm 150$ km/sec, $-267\pm 134$ km/sec, $-360\pm 141$ km/sec),
compared to the noise-free 1$\sigma$ sCDM expectation of 450 km/sec,
or the noise-free 1$\sigma$ tCDM expectation of 240 km/sec.

Neither the large variation of Hubble flows within $\sim 50h^{-1}$ Mpc
nor the large bulk flows may be surprising.
Figure 2(a) projects the positions of the inner 20 clusters (with a depth
of 56$h^{-1}$ Mpc) and the direction of their bulk motion on the sky.
The four clusters at the lower right quadrant of the map,
Telescopium at 24$h^{-1}$ Mpc, OC3627 at 29$h^{-1}$ Mpc, Pavo II at
37$h^{-1}$ Mpc and OC3742 at 40$h^{-1}$ Mpc, all show positive radial peculiar
velocities.  Since they lie roughly between us and the Great Attractor (GA,
at $\sim$4200 km/sec and in the direction of Hydro-Centaurus)
(Lynden-Bell et al. 1988), their positive radial peculiar velocities
may be the result of the gravitational
pull of the GA.  The clusters not in the general direction of the GA are
further away from it and thus experience a smaller infall into the GA.
As a result the subsample shows a large dipole in the direction of the four
clusters, plus a large positive monopolar flow.  Even the entire HM sample with
36 clusters going up to 11,000 km/sec may show the influence of the GA, 
since from figure 2(b), a projection of the outer 16 clusters, no new cluster
beyond the distance of the GA is added in the same direction of the four
clusters, and almost all the outer 16 clusters are in the opposite hemisphere.
To further show the effect of the four clusters, figure 3 plots $(H_2-H_1)/H_1$
vs. the depth $R$ of subsamples without the four clusters.  
The positive variation of Hubble flows at 40 to 60$h^{-1}$ Mpc becomes
much less significant and no longer exceeds the 2$\sigma$
expectations of the four chosen models. The bulk motion of the remaining
32 clusters, however, is still large, at
($718\pm 205$ km/sec, $-216\pm 140$ km/sec, $-344\pm 191$ km/sec).
Clearly, the HM sample is so
sparse and inhomogeneous that spurious bulk motions and peculiar
Hubble flows can result from structures such as the GA.  The window functions
$W^i({\bf k})$ of the sample, on the other hand, may not fully reflect
the density waves that give rise to these structures.  Therefore the
theoretical expectations based on these window functions can be too small.
This also implies that the assumption that the HM sample well
represents the true Hubble flow may rather be poor.

Similar calculations are done to the 277 galaxy groups of Mathewson et al.
(1992) plus the 11 clusters of Willick (1991) (M$+$W sample thereafter) 
in the Mark III catalogue.  The Aaronson et al. (1982) sample is too shallow
to be included.  The Mathewson et al. (1992) sample is much denser than
the HM sample.
It also goes up to $\sim 10,000$ km/sec, but only covers the southern
sky.  The Willick (1991) sample is too sparse, but it is not overly influenced 
by the GA (thanks to the large distances of its members from the GA).
Figure 4(a) and 4(b) show the projections of the two samples on the sky.  
Figure 5(a) shows the variation $(H_2-H_1)/H_1$ vs. the depth $R$ of
the subsamples of the M$+$W sample.
Figure 5(b) shows the comparison between this variation and 2$\sigma$
theoretical expectations (with noises included).
The errors and noise-included theoretical expectations are
calculated only for two representative points, because
both quantities are monotonic, slowly varying functions of $R$.
It should also be pointed out that $H_1$ is the Hubble expansion rate
of the HM sample because it provides the template Tully-Fisher relation
for the Mark III catalogue.  Therefore, instead of eq. (\ref{variation}),
\begin{equation}
\Bigl\langle\Bigl({H_2-H_1\over H_1}\Bigr)^2\Bigr\rangle
\approx
\Bigl\langle\Bigl({\delta H_1^{(v)}\over H_0}\Bigr)^2
+\Bigl({\delta H_2^{(v)}\over H_0}\Bigr)^2\Bigr\rangle
+\Bigl\langle\Bigl({\delta H_2^{(\epsilon)}\over H_0}\Bigr)^2\Bigr\rangle.
\end{equation}
The bulk motion of the M$+$W sample is relatively small,
($256\pm 57$ km/sec, $-343\pm 61$ km/sec, $170\pm 47$ km/sec), while the
noise-free expectation of sCDM is 670 km/sec, and that of tCDM is 361 km/sec.

There are two conclusions that can be drawn from figure 5: (1) the M$+$W sample
shows Hubble flows that are 6$\%$ to 8$\%$ faster than the Hubble flow of the
HM sample, with a $\sim 4\sigma$ significance; (2) the differences between the
HM sample of 36 clusters and the combined M$+$W sample are consistent with
theoretical expectations within 2$\sigma$, although the consistency is
marginal for models that predict smaller peculiar velocities, such as tCDM and
$\Lambda$CDM.  But because of the poor sampling of the HM sample, whether the
first conclusion implies a significant 
positive peculiar Hubble flow in the M$+$W sample is murky.

\section{Peculiar Hubble Flows in the LP Catalogue}

Lauer and Postman's Abell cluster catalogue is a volume-limited sample that
includes 119 Abell clusters with a redshift of $\la 0.05$ and
a galactic latitude above 13 degrees on both hemispheres.  Its advantages
over the HM sample are that it is volume-limited, more homogeneous, and deeper.
As table 1 shows, its Hubble flow is expected to deviate from the global
Hubble flow by only $\sim 1\%$.
The standard candle of the catalogue is taken
to be the luminosity $L$ of the brightest cluster galaxies as a function of
the second parameter $\alpha$, the power index of $L$ as a function of the
aperture. The distance to a BCG is taken to be its cosmological redshift
\footnote{The cosmological distortion is small and omitted.
See the discussion at the end of the section.}, i.e.,	
\begin{equation}
r_q=(cz_q-U_i\hat r_q^i-\delta H_1r_q)/H_0=(cz_q-U_i\hat r_q^i)/
(H_0+\delta H_1)=(cz_q-U_i\hat r_q^i)/H_1,
\label{distance}
\end{equation}
where $H_1$ is the Hubble expansion rate defined by the entire sample.
The estimated line-of-sight peculiar velocity is
\begin{equation}
S_q=cz10^{0.4[M_*(\alpha_q)-M_q]/(2-\alpha_q)},
\label{velocity}
\end{equation}
where $M_q$ is the absolute magnitude of BCG $q$, and $M_*(\alpha_q)$ is
the magnitude of the standard candle at $\alpha =\alpha_q$.

When applying eqs.~(\ref{likelihood}) to (\ref{Sq})
to the Lauer and Postman's catalogue,
because $M_q$ and $\alpha_q$ depend on $r_q$ and thus $U_i$ and
$\delta H$, one has to iterate eqs.~(\ref{likelihood}) to (\ref{Sq})
to obtain a self-consistent result.
Assuming no peculiar Hubble flow, Lauer and Postman (1994)
got a bulk flow of the sample
$U_x=477\pm 250$ km/sec, $U_y=-142\pm 273$ km/sec, $U_z=635\pm 198$ km/sec
relative to the rest frame of Cosmic Microwave Background Radiation.
I first repeat Lauer and Postman's calculation, log-linearly interpolating
table 3 of LP to find out the dependence of $M_q$ and $\alpha_q$ on $r_q$.
I get $U_x=475\pm 285$ km/sec, $U_y=-123\pm 308$ km/sec,
$U_z=648\pm 225$ km/sec, in very good agreement with LP's result.
The errors are estimated from eq.~(\ref{Unoise}), and verified by Monte
Carlo simulations similar to those of LP.

I then consider the possibility of a non-zero $\delta H_1=H_1-H_0$.
Since the distance scale is established within the sample itself,
$\delta H_1$ of the sample is really not a variable but a well-defined value.
In particular, one can always redefine the true Hubble expansion
rate as $H_0^\prime=H_0+\delta H_1=H_1$ so that variation of Hubble flows in
the sample can be investigated without referring to $H_0$, just like in
the previous section.

However, if one simply follows LP's way of calibrating $M_*(\alpha)$,
redefining $\delta H_1$ to zero cannot be done.  This is due to the fact
that the non-linear relation between
$S_q$ and $M_*(\alpha_q)-M_q$ always skews the distribution of
$S_q$ to the negative direction when $M_*(\alpha)-M_q$ is gaussian as
calibrated by regression from the
$M$-$\alpha$ distribution. An unphysical and negative
$\delta H_1/H_0$ will always result for such a calibration,
regardless of the value of $H_0$.  For the LP sample,
$\delta H_1/H_0\approx -1.2\%$.

One way to get rid of the unphysical $\delta H_1$ and set $\delta H_1$ to 0
is to calibrate $M_*(\alpha)$ by regression from the
$(S_q/cz_q)$-$\alpha$ distribution (my calculation shows that
the resulting $S_q/cz_q$ residual is gaussian with a standard
deviation of 0.166 at 6$\%$ C.L.).  The resultant bulk motion of the sample
is then $U_x=528\pm 285$ km/sec, $U_y=-272\pm 311$ km/sec, and
$U_z=607\pm 225$ km/sec.  $U_y$ being the most uncertain component,
shows the biggest difference from the LP result.  The resultant
$\delta H_1$ converges to zero to a high precision.

Once $M_*(\alpha)$ is calibrated, one can calculate the variation of 
Hubble flows within the sample. Figure 6 shows the variation with
$2\sigma$ error bars vs. the depth of the LP subsamples, and
its comparison with noise-included 2$\sigma$ theoretical expectations.
The error bars are once again calculated using
eq.~(\ref{variation}).  They are checked with
Monte Carlo simulations and consistency is found.
The figure shows evidence for negative $(H_2-H_1)/H_1$
within a radius of $\sim 60h^{-1}$ Mpc at the 2$\sigma$ level, indicating
a negative peculiar Hubble flow at this scale given the tiny deviation of
$H_1$ from $H_0$ (table 1).
But before the negative peculiar Hubble flow is explained with overdensities,
possible systematic effects have to be considered first, as in the case
of the Mark III catalogue.

Possible biases in the analysis have been discussed extensively
in LP (1994). Among them are several radially dependent biases that affect
the calculation of peculiar Hubble flows. The first is the selection bias.
A luminosity-limited catalogue
introduces an artificial Hubble outflow due to the
missing low brightness galaxies.
This is apparently not the case with LP's catalogue
because it is volume-limited.
Secondly, the estimated peculiar velocity depends on the deceleration
parameter $q_0$ assumed. But for a catalogue extending only to $z\approx 0.05$,
the estimated radial peculiar velocity
is only changed by $\la 1\%$ if the true $q_0$ is changed by 0.5.
Since the standard candle in LP's catalogue is established internally,
influenced mostly by outlying clusters, the effect on the
peculiar Hubble flow of inner clusters is $\la 1\%$.
A third bias comes from the random peculiar velocities of BCGs due to local
non-linearities which tend to scatter more BCGs to lower measured
redshifts than to higher redshifts. But given a typical value
of this random radial velocity of 300 km/sec, the velocity bias introduced
on the 6000 km/sec scale is only 0.5$\%$ (LP).
Another important concern is whether the BCGs of the inner Abell clusters
belong statistically to the same population of the entire BCG
sample.  Table 2 lists the statistical properties of the inner BCG subsamples
and the entire BCG sample. They are consistent statistically.

But a problem arises when one tests whether the detection
is dominated by a small number of clusters.  Figure
7 projects the inner 11 clusters (which show the negative
peculiar Hubble flow at 2$\sigma$) on the sky.  The subsample is very
inhomogeneous, due to its small size.  In particular, there is one (and only
one) cluster (A262) that roughly aligns with the bulk motion of the subsample.
Since the bulk motion and the peculiar Hubble flow are not determined
independently, the cluster will certainly play a dominate role in determining
both quantities.  If the clusters is excluded, as shown in the small window of
figure 6(a), no variation of Hubble flows is found at the
2$\sigma$ level.  Therefore, once again, the poor sampling of the subsample
renders its result susceptible to systematic effects such as small scale
structures, unrepresented structures, small number statistics, etc.

\section{Peculiar Hubble Flows in a Type Ia supernova sample}

Type Ia supernova samples probe much deeper ($\sim 10^3$ Mpc) and have
very precise distance measurements ($\sim 5\%$).
Therefore, even for the 20 Type Ia SNe compiled by Riess et al. (1996),
table 1 shows that the expected 1$\sigma$ deviation of its Hubble 
flow from the global one is only $\sim 1\%$.
But because of its extremely sparse sampling, its usefulness to investigations
of peculiar Hubble flows is significantly compromised.  
Figure 8 shows the variation of Hubble flows in the Type Ia SN sample.
No significant variation is found at any scale because of its sparse
sampling.  Since the errors in the plot roughly scale as the inverse
square root of the number of objects, as more Type Ia
supernovae are being observed with good time coverage, they can certainly
provide more accurate results at a wide range of depth in the future.

\section{Discussion}

After investigating the variation of Hubble flows in the Mark III catalogue
(Willick et al. 1996), the Lauer and Postman (1994) sample, and the Type Ia
supernova sample of Riess et al. (1996),
there are three conclusions that I would like to draw:

\begin{itemize}
\item
Some significant peculiar Hubble flows are found at face value in the
Mark III catalogue and at 50 to 60$h^{-1}$ Mpc in the Lauer and Postman
sample.  They at face value disfavor cosmological models predicting
smaller peculiar velocities, such as tCDM.  However, because of the sparse
and inhomogeneous sampling of the HM cluster sample on which the
Hubble expansion of the Mark III catalogue is defined, and
of the sparse sampling of the Lauer and Postman sample within the 60$h^{-1}$
Mpc scale, the peculiar Hubble flows found are dominated by statistics
of a small number of clusters, and thus cannot be trusted at their face value.
Further supporting this is the fact that the peculiar Hubble flows found in the
different samples are inconsistent: the HM sample shows positive peculiar
Hubble flows below $\sim 70h^{-1}$ Mpc, while the M$+$W sample shows positive
peculiar Hubble flows up to $\sim 100h^{-1}$ Mpc; the LP sample, on the
other hand, shows negative peculiar Hubble flows at $\sim 60h^{-1}$ Mpc.

\item 
All samples agree on that the Hubble flows beyond a depth of
$\sim 80h^{-1}$ Mpc are true to the global Hubble flow to better than 10$\%$
at $\sim 2\sigma$ C.L.  In each sample, the
variation of Hubble flows at a depth of $\geq 80h^{-1}$ Mpc,
with respect to the Hubble flow of entire sample, are found to be
$<10\%$ at 2$\sigma$.  This combined with the small expected deviation 
(1 to 3$\%$ at 1$\sigma$, see table 1)
of the Hubble flow defined by each sample from the global Hubble flow, ensures
that the Hubble flows at $\geq 80h^{-1}$ Mpc conform to the global Hubble flow
to better than 10$\%$ at $\sim 2\sigma$ C.L.
At depth below 80$h^{-1}$ Mpc, the limits on peculiar Hubble flows are
somewhat weaker.  For example, at the depth of the Coma cluster (70$h^{-1}$
Mpc), while the Lauer and Postman sample indicates a less than
15$\%$ deviation from the true Hubble flow at 2$\sigma$, all others
imply a stronger limit--less than 10$\%$ deviation from the true Hubble flow
at 2$\sigma$ C.L.

\item
Since the peculiar Hubble flows detected in the samples are not trustworthy
at the moment, so is its comparison with theoretical predictions.
However, this is not to say that any such comparison will be
useless, because the major limiting factor to a meaningful comparison between
models and observations is neither our uncertain knowledge of
the Hubble constant, nor the uncertainties in the distance scales,
but rather the imperfect sampling of our local universe.
The errors achieved by the current Mark III catalogue are comparable
($\sim 1\%$ at $\sim 60h^{-1}$ Mpc) to the differences in noise-free model
predictions.  Therefore, If samples were
improved to yield a consistent and reliable detections of peculiar Hubble
flows, measurements of peculiar Hubble flows would certainly
be able to test models.
\end{itemize}

While the first and second conclusions represent an attempt
to map out Hubble flows in our local universe, which are extremely
interesting in their own rights, the third conclusion aims
at finding a potential tool to gain further information on cosmological
parameters, and therefore warrants a further discussion.
Since errors in the question are roughly proportional to the square root of
the number of objects in a sample, if a deep ($>100h^{-1}$ Mpc), homogeneous
(covering all directions and depth)
and dense (at least 10$^2$ clusters to yield a template Tully-Fisher
relation and $\sim 10^3$ clusters and groups within 100$h^{-1}$
Mpc to be sampled) sample can be assembled with the Tully-Fisher relation,
a reliable detection of peculiar Hubble flows may be made with
errors at the $\la 1\%$ level, sufficient to test many models at the
$\sim 60h^{-1}$ Mpc scale.  Or, if more than 100
Type Ia supernovae are adequately observed as distance indicators, the
errors of their measured peculiar Hubble flows can be cut by more than 1/2 from
the current level to $\sim 1\%$.  Better yet, if the Tully-Fisher relation
is calibrated with Type Ia supernovae, by observations of Type Ia
supernovae in clusters, not only the overall Hubble flow of the Tully-Fisher
sample will represent the global Hubble flow better, the two samples can be
combined to yield better statistics.

\section{Acknowledgement}

The author thanks A. Dekel, M. Postman, L. Widrow, J. Willick and the
referee for helpful discussions and suggestions,
and thanks Clarence Lee for proofreading the manuscript.
The work is supported by grants NASA NAG5-3062 and NSF PHY95-03384 at UCSD.

\begin{deluxetable}{ccccccc}
\footnotesize
\tablecaption{Parameters of four chosen models\tablenotemark{*}\  and their
predicted 1$\sigma$ deviations from a global Hubble flow for the Han \& Mould
sample (HM), the Lauer \& Postman sample (LP), and the Type Ia supernova
sample (SN).}
\tablewidth{0pt}
\tablehead{\colhead{} && \colhead{sCDM} & \colhead{tCDM} 
& \colhead{$\Lambda$CDM} & \colhead{CHDM}}

\startdata
$\Omega_{\rm tot}$& &1   &1   &1   &1   \nl
$\Omega_m$& &1   &1   &0.3 &1   \nl
$\Omega_b$& &0.06&0.06&0.03&0.06\nl
$\Omega_\nu$&&0  &0   &0   &0.2 \nl
$h$       & &0.5 &0.5 &0.7 &0.5 \nl
$n$       & &1   &0.7 &1   &0.9 \nl
	  & &    &    &    &    \nl
$\Bigl\langle\Bigl({H_1-H_0\over H_0}\Bigr)^2\Bigr\rangle^{1/2}_{\rm HM}$
& & 3.1$\%$ & 1.8$\%$ & 2.1$\%$ & 2.4$\%$ \nl
$\Bigl\langle\Bigl({H_1-H_0\over H_0}\Bigr)^2\Bigr\rangle^{1/2}_{\rm LP}$
& & 1.2$\%$ & 0.7$\%$ & 0.9$\%$ & 0.9$\%$ \nl
$\Bigl\langle\Bigl({H_1-H_0\over H_0}\Bigr)^2\Bigr\rangle^{1/2}_{\rm SN}$
& & 2.0$\%$ & 1.0$\%$ & 1.1$\%$ & 1.3$\%$ \nl
\enddata
\tablenotetext{*}{The power spectra of the models are from Bardeen et al.
(1986), Bunn and White (1996), Ma (1996) and Sugiyama (1995).}
\end{deluxetable}

\begin{deluxetable}{ccccc}
\footnotesize
\tablecaption{Comparison of statistical properties in the LP sample.}
\tablewidth{300pt}
\tablehead{\colhead{N}\tablenotemark{a}
& \colhead{$\langle\alpha\rangle$}\tablenotemark{b}
& \colhead{$\langle M_q-M_*(\alpha_q)\rangle$}\tablenotemark{c}
& \colhead{$\sigma_{M_q-M_*(\alpha_q)}$}   \tablenotemark{d}
& \colhead{$P_{KS}$}      \tablenotemark{e}}

\startdata
8  & 0.61 & 0.141 & 0.255 & 83$\%$ \nl
9  & 0.61 & 0.113 & 0.253 & 71$\%$ \nl
10 & 0.59 & 0.109 & 0.239 & 84$\%$ \nl
11 & 0.59 & 0.106 & 0.227 & 93$\%$ \nl
   &      &       &       &        \nl
119& 0.57 & 0     & 0.245 & 39$\%$ \nl

\enddata
\tablenotetext{a}{The number of clusters in the (sub)sample.}
\tablenotetext{b}{The average $\alpha$ in the (sub)sample.}
\tablenotetext{c}{The average $M_q-M_*(\alpha_q)$ in the (sub)sample.}
\tablenotetext{d}{The standard deviation of $M_q-M_*(\alpha_q)$.}
\tablenotetext{e}{The confidence level that $M_q-M_*(\alpha_q)$ is
gaussian according to a K-S test.}
\end{deluxetable}

\clearpage

\figcaption[fig1.ps]{(a) The variation of Hubble flows in the HM
sample with $2\sigma$ error bars, as a function of the depth of subsamples.
(b) The crosses show the variation of Hubble flows in the HM sample.
Curves are the 2$\sigma$ expectations (with noises included) from models.
Model expectations for subsamples
are connected for a simpler presentation. \label{fig1}}

\figcaption[fig2.ps]{(a) Projection of the inner 20 clusters of the HM sample
and their bulk motion in galactic coordinates.
Solid circles denote clusters with negative
radial peculiar velocities, open circles denote clusters with positive
radial peculiar velocities.  The cross is the direction of the bulk motion.
(b) Projection of the outer 16 clusters of
the HM sample in galactic coordinates.  The dotted line is the equator of
the sky when the bulk motion of the inner 20 clusters is chosen as a pole.
\label{fig2}}

\figcaption[fig3.ps]{The variation of Hubble flows in the HM
sample less four clusters in the lower right quadrant of figure 2(a),
with $2\sigma$ error bars.\label{fig3}}

\figcaption[fig4.ps]{(a) Projection of 277 groups of the Mathewson et al.
(1992) sample in galactic coordinates. (b) Projection of 11 clusters of the
Willick (1991) sample in galactic coordinates.  Legends
are the same as in figure 2. \label{fig4}}

\figcaption[fig5.ps]{Similar to figure 1 for the M$+$W sample.  Errors and
theoretical expectations are calculated only for one subsample and the
entire sample.
\label{fig5}}

\figcaption[fig6.ps]{Similar to figure 1 for the Lauer and Postman (1994)
sample.  Plot inside the window of (a) shows the variation of Hubble flows
when cluster A262 is excluded. \label{fig6}}

\figcaption[fig7.ps]{Projection of the inner 11 Abell clusters in the
Lauer and Postman sample and their bulk motion in galactic coordinates.
Legends are the same as in figure 2. \label{fig7}}

\figcaption[fig8.ps]{Similar to figure 1(a) for the sample of
20 Type Ia supernovae.\label{fig8}}

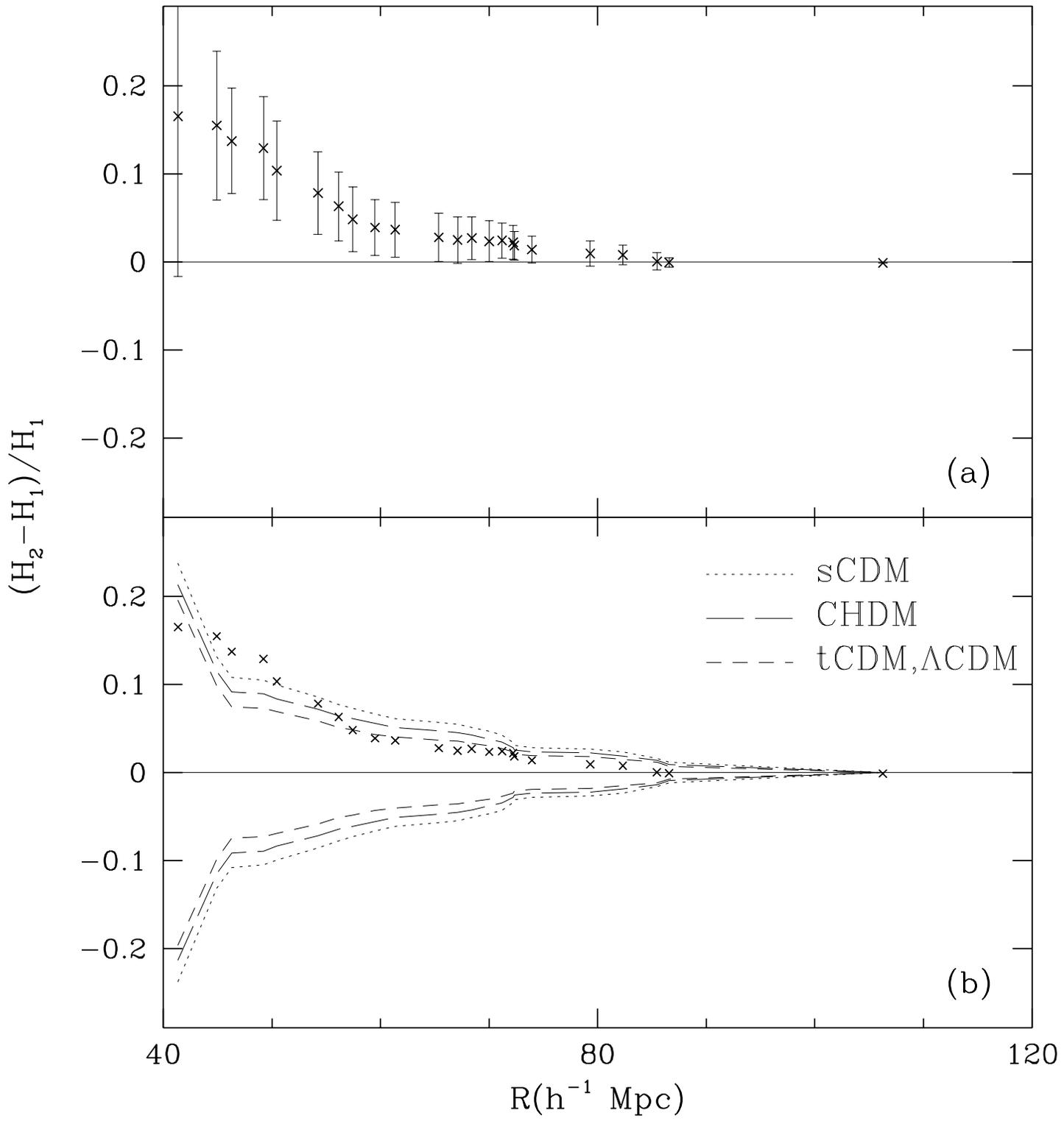
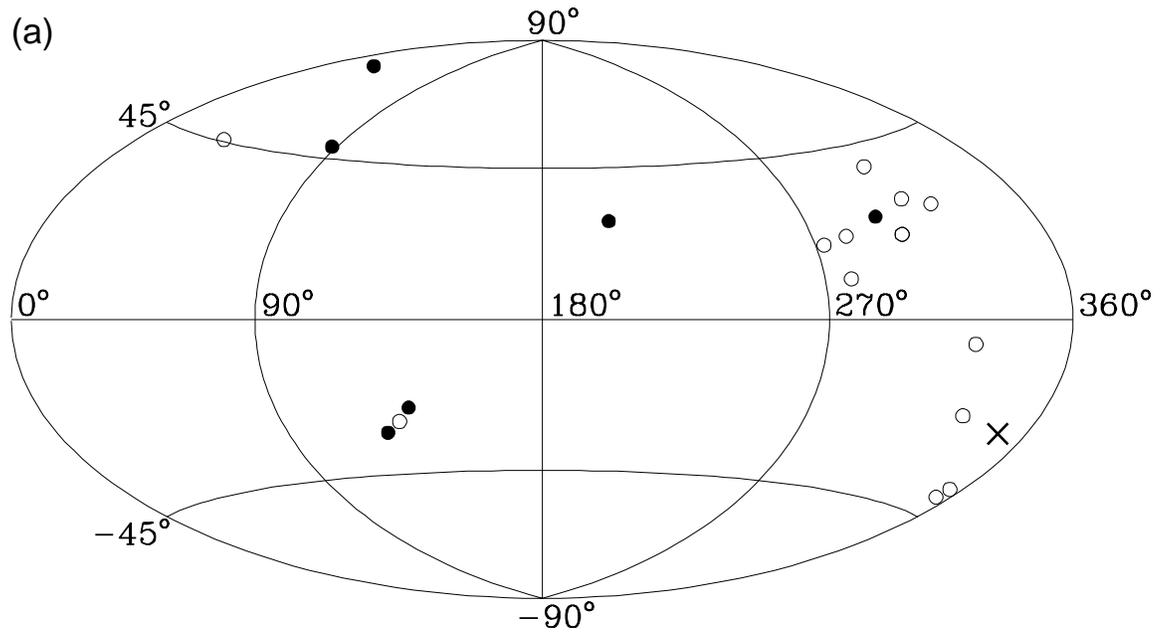
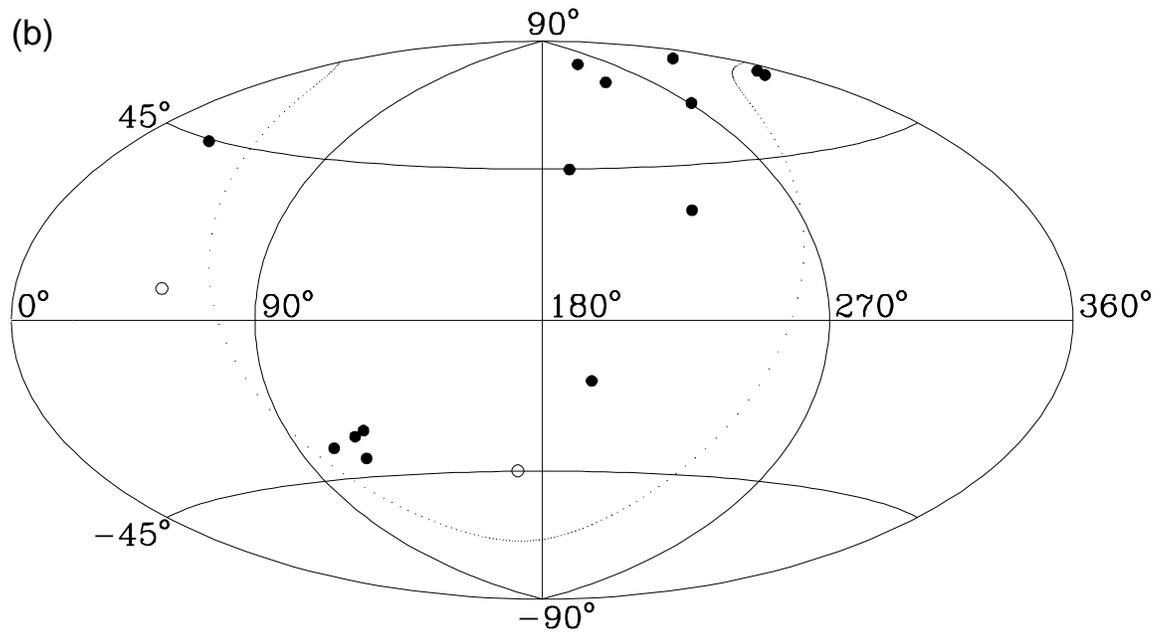
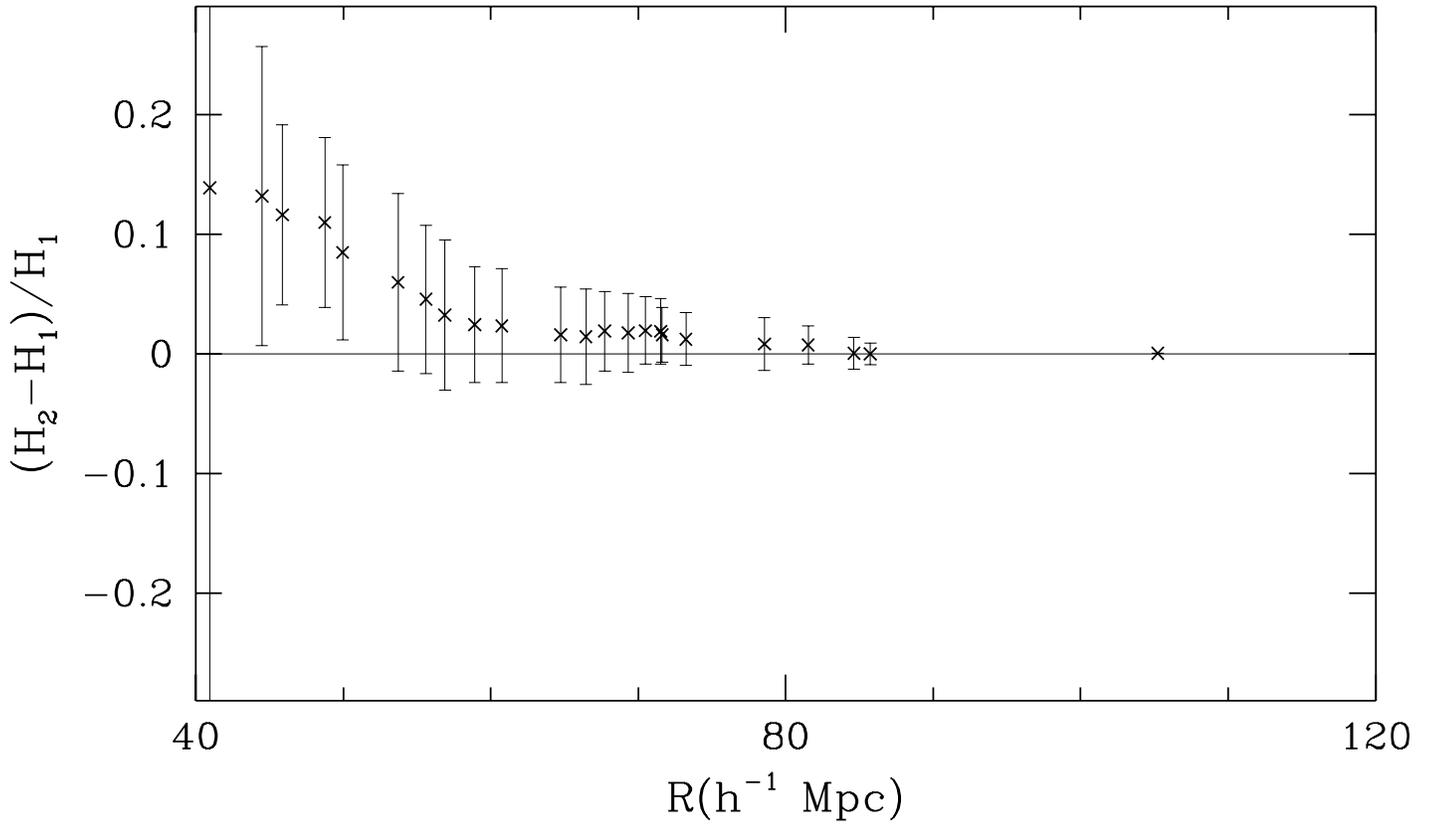
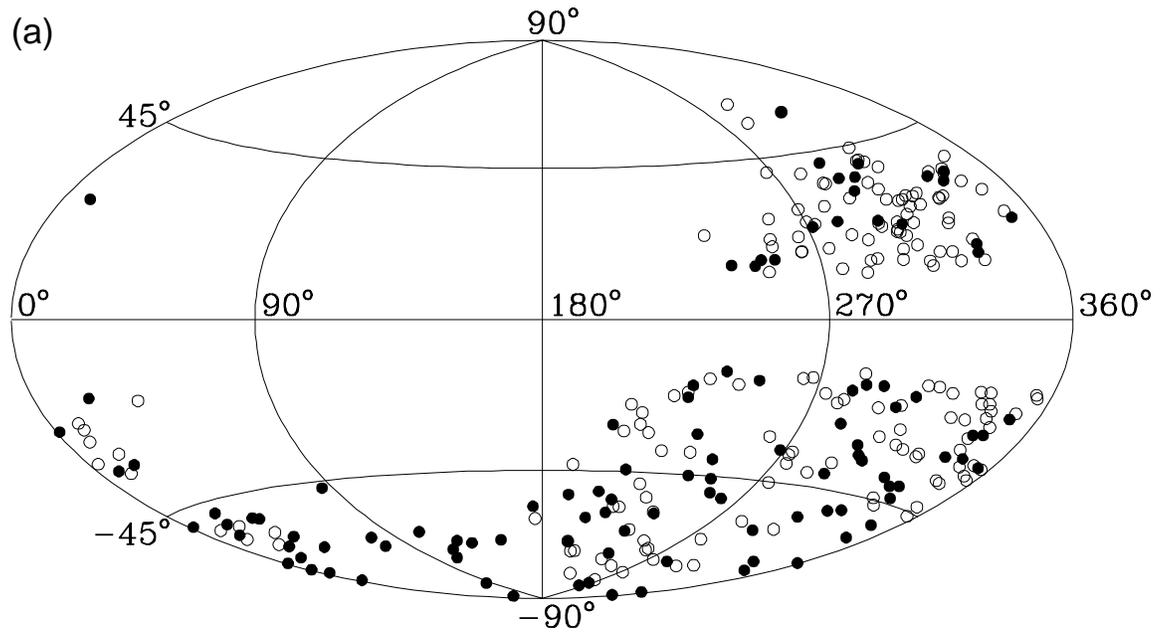
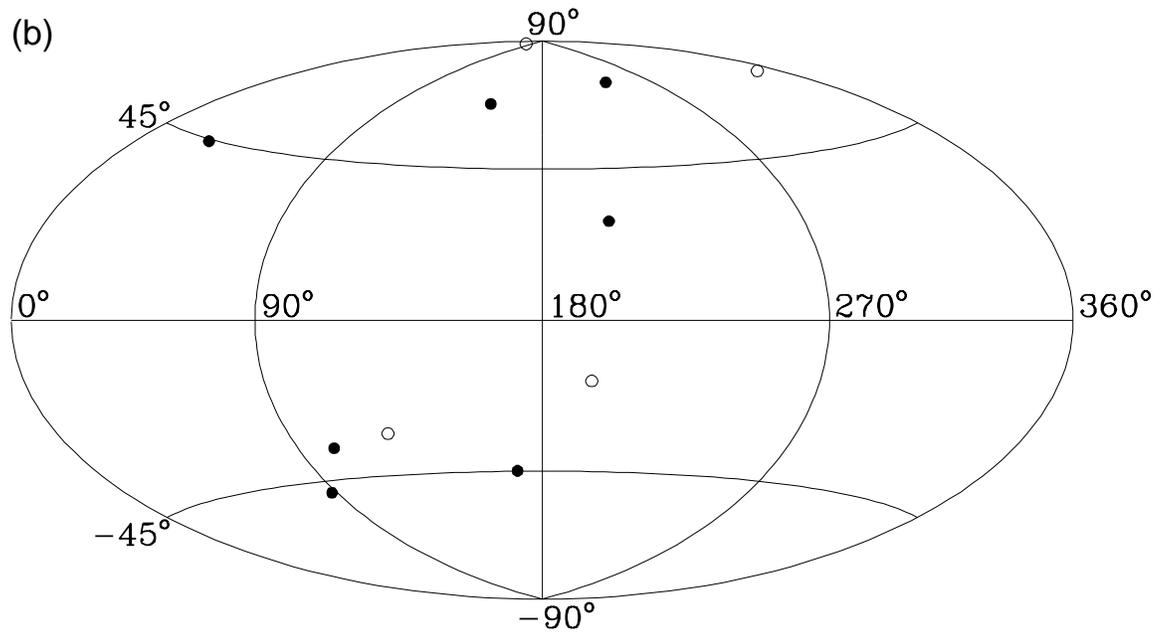
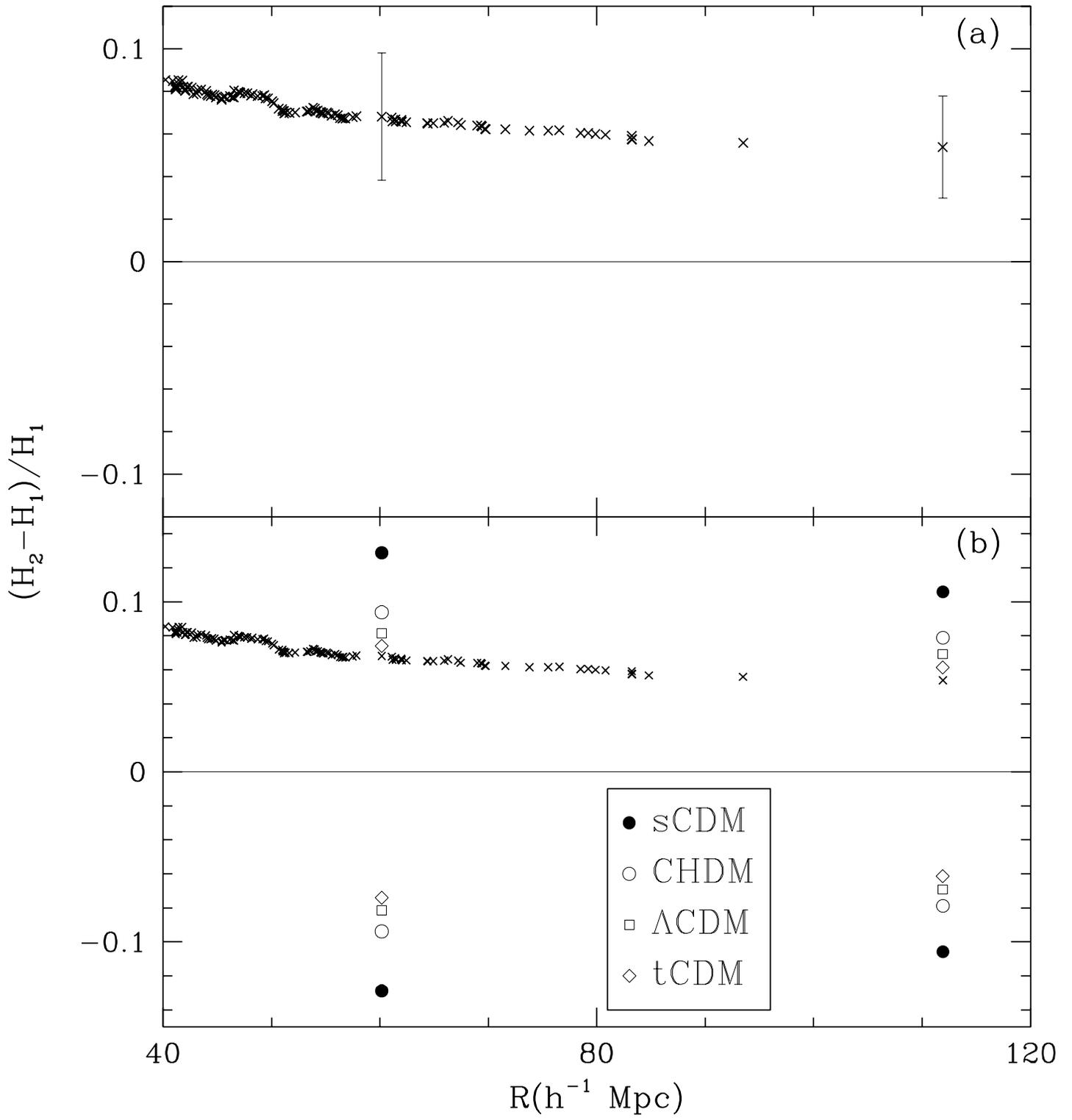
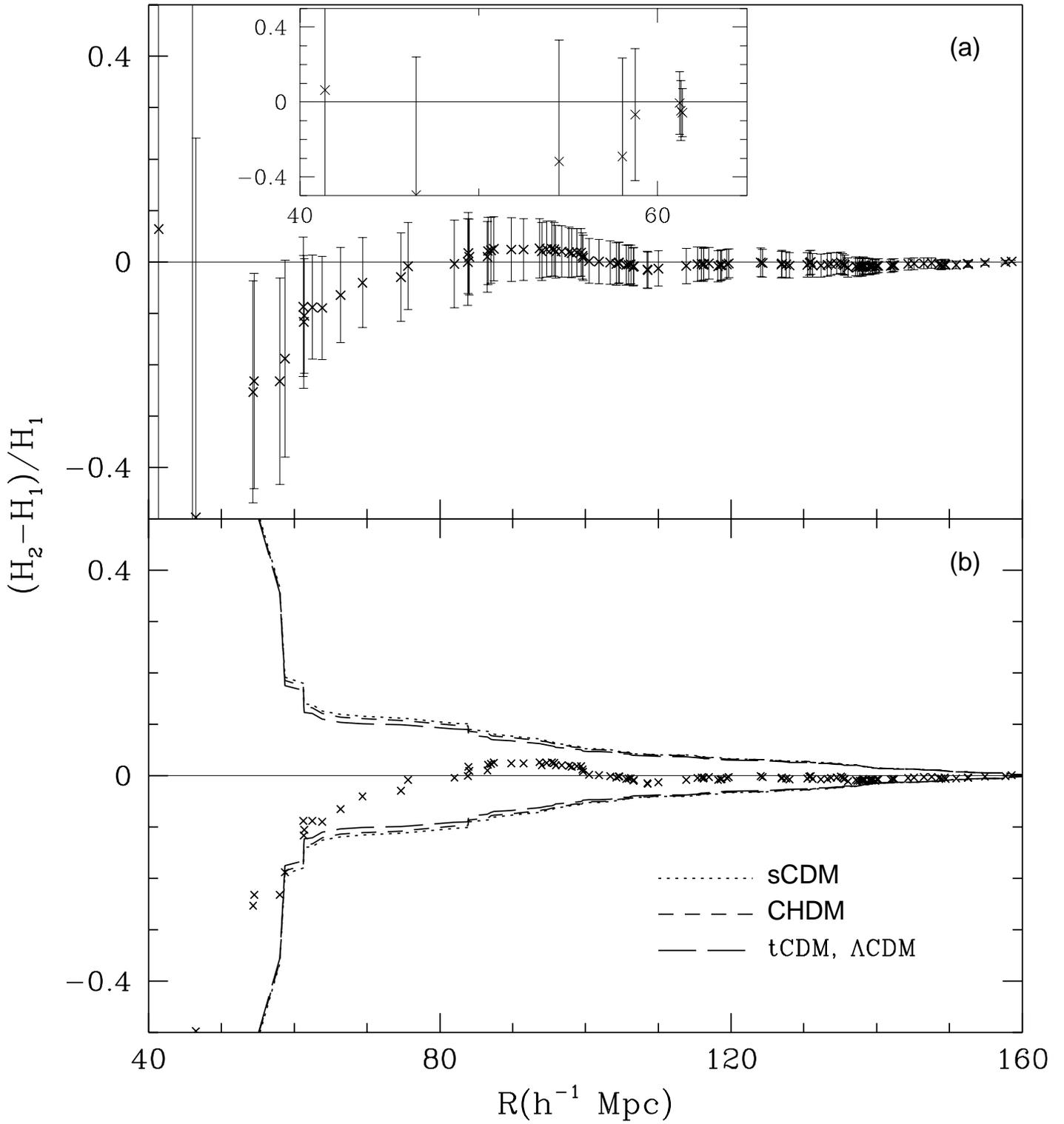
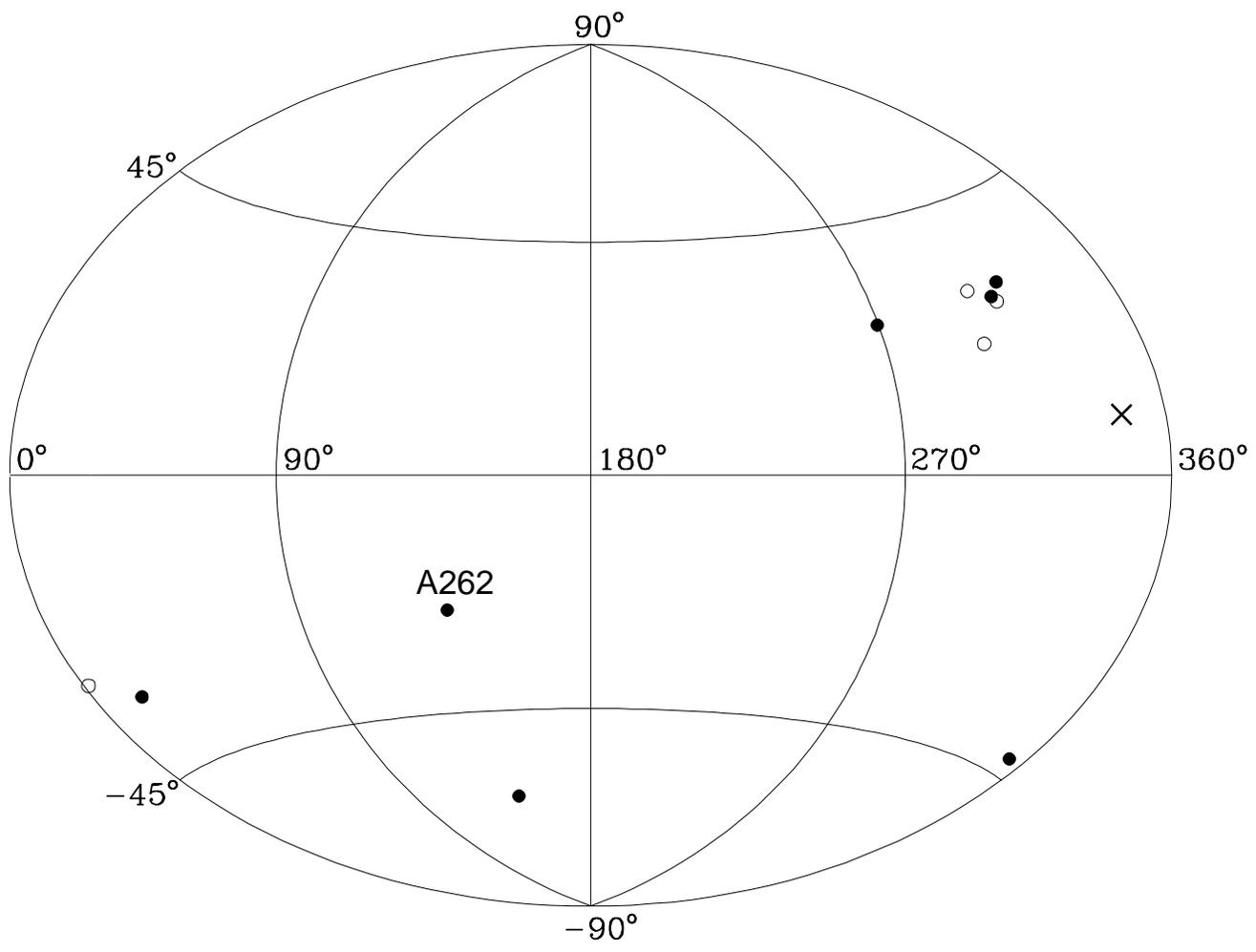
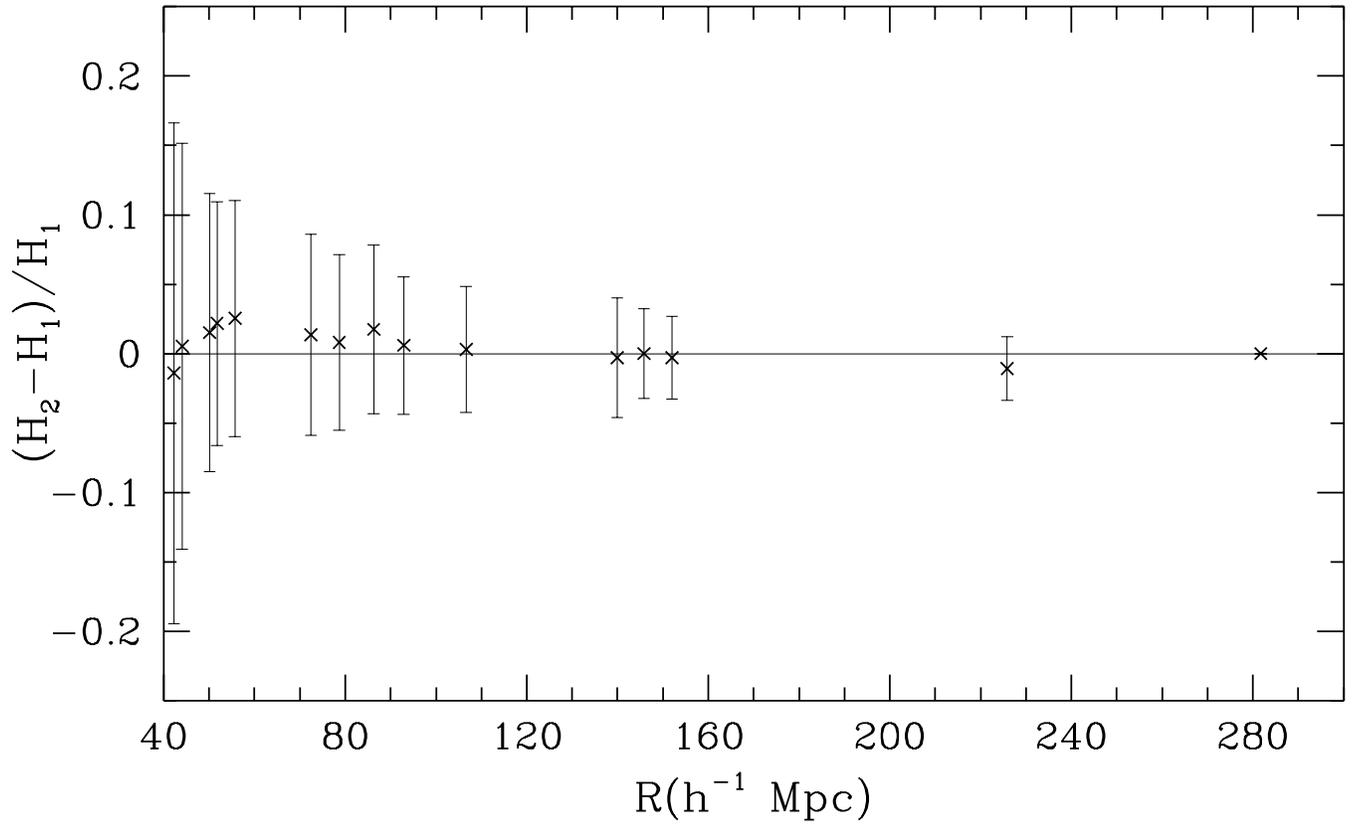
\end{document}